\documentclass[amsmath,amssymb,pra,final,showpacs,twocolumn]{revtex4-1}

\usepackage{bm,hyphenat,xspace}
\usepackage{graphicx,epsfig}

\newcommand {\mbf}[1]{{\mathbf{#1}}}

\newcommand {\mcu}{\mathcal{U}}

\begin{document}

\title {Momentum-space calculation of four-boson recombination}

\author{A.~Deltuva} 
\affiliation{Centro de F\'{\i}sica Nuclear da Universidade de Lisboa, 
P-1649-003 Lisboa, Portugal }

\received{December 12, 2011}

\begin{abstract}The system of four identical bosons with large two-boson 
scattering length is described using momentum-space 
integral equations for the four-particle transition operators. 
The creation of Efimov trimers via ultracold
 four-boson recombination is studied. The universal behavior
of the  recombination rate is demonstrated.
\end{abstract}

 \maketitle

\section{Introduction}

In 1970 V. Efimov predicted that with vanishing  two-particle binding
 an infinite number of weakly bound states 
with total orbital angular momentum $\mathcal{L}=0$ may exist 
in the three-particle system \cite{efimov:plb,braaten:rev}.
Such a situation takes place if at least two pairs have
infinite scattering length $a \to \infty$ \cite{braaten:rev}.
Away from this limit the number of trimers is finite but,
depending on $a$, may be large \cite{efimov:plb,braaten:rev}.
This three-particle Efimov effect has an impact on the behavior
of more complex few-particle systems. In particular,
for each Efimov trimer the existence of two tetramers 
 was predicted in Refs.~\cite{hammer:07a,stecher:09a}.
The signatures of these states were observed recently in 
experiments with ultracold atoms where
the formation of the tetramers led to a resonant enhancement 
of the recombination or relaxation processes
\cite{ferlaino:08a,ferlaino:09a,pollack:09a}.

These first steps in exploring the four-body Efimov physics experimentally
also call for accurate theoretical studies of the four-body systems with 
large $a$. Following our recent works devoted to the bosonic atom-trimer 
\cite{deltuva:10c} and dimer-dimer \cite{deltuva:11b} scattering, 
here we focus on the four-atom recombination process
in the system of four identical bosons.
The description is based on the exact four-particle equations
 for the transition operators  \cite{grassberger:67}
derived by Alt, Grassberger, and Sandhas (AGS).
For the recombination process we need the  transition operator
connecting two- and four-cluster channel states; its relation
to the two-cluster AGS operators calculated in Refs.
\cite{deltuva:10c,deltuva:11b,deltuva:07a,deltuva:07c}, 
will be given in the present work. 
We solve the integral AGS equations using momentum-space framework.
In previous works we demonstrated its reliability for reactions
 involving many very weakly bound trimers as it happens in the universal 
regime where the accuracy  of the coordinate-space methods 
may be more limited \cite{stecher:09a,lazauskas:he,deltuva:ef}.
For example, in Ref.~\cite{deltuva:11a} we predicted universal
intersections of shallow tetramers with the corresponding atom-trimer 
thresholds while the adiabatic hyperspherical 
calculations of Refs.~\cite{ferlaino:09a,stecher:09a} were not 
sufficiently precise to find this remarkable feature that
 leads to a resonant behavior in the ultracold atom-trimer collisions.
Thus, we expect that also in the case of the four-boson recombination 
we will be able to achieve the universal limit with higher accuracy
compared to the existing coordinate-space calculations
\cite{stecher:09a,mehta:09a}.
We note that approximate semi-analytical recombination results have been
obtained in Ref.~\cite{PhysRevLett.102.133201} for a system of 
three identical bosons plus a distinguishable particle.

An extension of the four-body scattering calculations above the 
four-cluster breakup threshold is a very serious theoretical challenge
both in atomic and nuclear physics. In the momentum-space framework
the difficulties arise due to complicated singularity structure 
in the kernel of the integral equations. Not all but some of these
difficulties are present already at the four-cluster breakup threshold.
In this work we  restrict ourselves to a latter case which is
sufficient to calculate the  four-atom recombination
in the ultracold limit.
Nevertheless, the present work is an important intermediate step 
towards solving the momentum-space four-body scattering equations 
at positive energies.

We use a system of units where $\hbar=1$ and therefore is omitted
in the equations unless needed for dimensional analysis.
In Sec.~\ref{sec:4bse} we recall the AGS equations and derive the relation
between the two- and four-cluster transition operators.
In Sec.~\ref{sec:res} we present results for the 
four-boson recombination.
We summarize in  Sec.~\ref{sec:sum}.

\section{Four-boson scattering \label{sec:4bse}}

We consider a system of four spinless particles with Hamiltonian 
\begin{equation}
H = H_0 + \sum_{i=1}^6 v_i ,
\end{equation}
 where $H_0$ is the kinetic energy
operator for the relative motion and $v_i$ the short-range potential
acting within the pair $i$; there are six pairs.
For the description of the scattering process in such a system
we use the AGS equations \cite{grassberger:67}.  They are
exact quantum-mechanical equations of the Faddeev-Yakubovsky (FY) type.
However, in contrast to the original FY equations \cite{yakubovsky:67}
 for the wave-function components, the AGS equations
are formulated for the transition operators in the integral form
and therefore are better suited to be solved in the momentum-space framework
preferred by us. 

We aim to determine the amplitude for the recombination of
four free particles into a two-cluster state. Due to time reversal
symmetry it is equal to the  amplitude for the four-cluster
breakup of the initial two-cluster state. The latter is more
directly related to the AGS transition operators calculated
in our previous works \cite{deltuva:10c,deltuva:11b,deltuva:07a,deltuva:07c}.

\subsection{AGS equations \label{sec:ags}}

In a compact notation the AGS equations can be written as 18-component
matrix equations \cite{grassberger:67,deltuva:ef}.
The components are distinguished by the chains of partitions, i.e.,
by the two-cluster partition 
and by the three-cluster partition. Obviously, the two-cluster partitions
may be of $3+1$ or $2+2$ type; we will denote them by Greek subscripts.
All  three-cluster partitions are of $2+1+1$ type and are specified
by the pair of particles that we will denote by Latin superscripts.
For our consideration we need the explicit form of the
AGS equations for the transition operators, i.e.,
\begin{equation} \label{eq:AGSgen}
\mcu_{\sigma \rho}^{ji} = 
(G_0\, t_i\, G_0)^{-1}\,
\bar{\delta}_{\sigma \rho} \, \delta_{ji} 
  + \sum_{\gamma k} \bar{\delta}_{\sigma \gamma}  U_{\gamma}^{jk} G_0\, t_k\, 
G_0 \, \mcu_{\gamma \rho}^{ki}.
\end{equation}
Here $\bar{\delta}_{\sigma \rho} = 1 - {\delta}_{\sigma \rho}$,
\begin{equation} \label{eq:G0}
G_0 = (E+i0-H_0)^{-1}
\end{equation}
 is the free resolvent with the available
four-particle energy $E$,
\begin{equation} \label{eq:t2b}
t_i = v_i + v_i G_0 t_i
\end{equation}
is the two-particle transition matrix for the pair $i$, and
\begin{gather}
U_{\gamma}^{jk} = G^{-1}_0 \, \bar{\delta}_{jk} + \sum_i \bar{\delta}_{ji}
\, t_i \, G_0 \, U_{\gamma}^{ik}
\end{gather}
is the subsystem transition operator of $3+1$ or $2+2$ type, 
depending on $\gamma$. In the summations over the pairs only the ones
 internal to the respective two-cluster partition contribute.

The transition operators $\mcu_{\sigma \rho}^{ji}$ contain full information
on the scattering process. Their  on-shell matrix elements
$ \sum_{ji} \langle \phi_{\sigma,n'}^j |  \mcu_{\sigma \rho}^{ji} 
| \phi_{\rho,n}^i \rangle $ are the amplitudes 
$\langle \Phi_{\sigma,n'}| T_{\sigma \rho} | \Phi_{\rho,n} \rangle $
for the two-cluster reactions. Here
\begin{equation} \label{eq:phi}
| \phi_{\rho,n}^i \rangle = G_0 \sum_{j} \bar{\delta}_{ij} t_j 
| \phi_{\rho,n}^j \rangle
\end{equation}
are the Faddeev components of the initial and final channel states
\begin{equation} \label{eq:Phi}
| \Phi_{\rho,n} \rangle =  \sum_{i} | \phi_{\rho,n}^i \rangle,
\end{equation}
and $n$ distinguishes between the different channel states in the same 
partition.  $| \Phi_{\rho,n} \rangle$ is given by the
$n$th bound state wave function in partition $\rho$  times the plane 
wave with momentum  $\mathbf{p}_{\rho,n}$ between the clusters.
and is normalized to 
$\langle \Phi_{\sigma,n'} | \Phi_{\rho,n} \rangle = \delta_{\sigma \rho} \,
\delta_{n'n} \, \delta(\mathbf{p}_{\sigma,n'}-\mathbf{p}_{\rho,n})$.
The on-shell condition for each channel state relates
its binding energy, relative two-cluster momentum and reduced mass as
$ \epsilon_{\rho,n} + p_{\rho,n}^2/2\mu_\rho  = E$.

The full wave function corresponding to the initial
two-cluster channel state $| \Phi_{\rho,n} \rangle$
is also determined by the AGS transition operators, i.e.,
\begin{equation} \label{eq:AGSpsi}
| \Psi_{\rho,n} \rangle = 
| \Phi_{\rho,n} \rangle  
  + \sum_{\gamma jki} 
G_0 \, t_j \, G_0 \,  U_{\gamma}^{jk} G_0\, t_k\, G_0 \, 
\mcu_{\gamma \rho}^{ki} | \phi_{\rho,n}^i \rangle.
\end{equation}

The amplitude for the four-cluster breakup reaction can be obtained as
\begin{equation} \label{eq:T0}
 \langle \Phi_{0} |  T_{0 \rho} | \Phi_{\rho,n} \rangle 
= \langle \Phi_{0} | \sum_{i=1}^{6} v_i | \Psi_{\rho,n} \rangle .
\end{equation}
The wave function \eqref{eq:AGSpsi} satisfies also the
Schr\"odinger equation, thus, 
$ \sum_{i=1}^{6} v_i | \Psi_{\rho,n} \rangle = G_0^{-1} | \Psi_{\rho,n} \rangle$.
Furthermore, $ G_0^{-1} | \Phi_{0} \rangle = 0$ since 
the four-cluster channel state $| \Phi_{0} \rangle$
is an eigenstate of $H_0$ with eigenvalue $E$.
In fact,  $| \Phi_{0} \rangle$ is a product of three plane waves
(each is normalized to the Dirac $\delta$-function)
corresponding to the relative motion of four free particles.
Thus, the amplitude for the  four-cluster breakup of the two-cluster
initial state is
\begin{equation} \label{eq:AGS0}
 \langle \Phi_{0} |  T_{0 \rho} | \Phi_{\rho,n} \rangle
=  \sum_{\gamma jki} \langle \Phi_{0} |
t_j \, G_0 \,  U_{\gamma}^{jk} G_0\, t_k\, G_0 \, 
\mcu_{\gamma \rho}^{ki} | \phi_{\rho,n}^i \rangle.
\end{equation}
Due to time reversal symmetry it describes also the
four-particle recombination into a two-cluster state, i.e.,
$  \langle \Phi_{\rho,n} |  T_{\rho 0} | \Phi_{0} \rangle
= \langle \Phi_{0} |  T_{0 \rho} | \Phi_{\rho,n} \rangle $.

If all four particles are identical, there are only 
two distinct two-cluster partitions, one of $3+1$ type and
one of $2+2$ type. 
 We choose those partitions to be ((12)3)4 and
(12)(34) and denote them in the following by $\alpha =1$ and $2$,
respectively. The AGS equations \eqref{eq:AGSgen} for the
symmetrized transition operators $\mcu_{\beta\alpha}$ become
\begin{subequations} \label{eq:U}
\begin{align}  
\mcu_{11}  = {}&  P_{34} (G_0  t  G_0)^{-1}  
 + P_{34}  U_1 G_0  t G_0  \mcu_{11} + U_2 G_0  t G_0  \mcu_{21} , 
\label{eq:U11} \\  
\mcu_{21}  = {}&  (1 + P_{34}) (G_0  t  G_0)^{-1}  
+ (1 + P_{34}) U_1 G_0  t  G_0  \mcu_{11} , \label{eq:U21} \\
\mcu_{12}  = {}&  (G_0  t  G_0)^{-1}  
 + P_{34}  U_1 G_0  t G_0  \mcu_{12} + U_2 G_0  t G_0  \mcu_{22} , 
\label{eq:U12} \\  
\mcu_{22}  = {}& (1 + P_{34}) U_1 G_0  t  G_0  \mcu_{12} . \label{eq:U22}
\end{align}
\end{subequations}
Here the two-particle transition matrix $t$ acts within the pair (12)
 and the symmetrized operators for the $3+1$ and $2+2$ subsystems
are obtained from the integral equations
\begin{equation} \label{eq:U3}
U_{\alpha} =  P_\alpha G_0^{-1} + P_\alpha  t G_0  U_{\alpha}.
\end{equation}
 The employed basis states have to be symmetric 
under exchange of two particles in subsystem (12) for $3+1$ partition
and in (12) and (34) for $2+2$ partition. 
The correct symmetry of the four-boson system is ensured by the 
operators $P_{34}$, $P_1 =  P_{12}\, P_{23} + P_{13}\, P_{23}$, and
$P_2 =  P_{13}\, P_{24} $ where $P_{ab}$ is the
permutation operator of particles $a$ and $b$.
For each two-cluster channel state  
$| \Phi_{\alpha,n} \rangle = (1+ P_\alpha)  |\phi_{\alpha,n} \rangle$
there is only one independent Faddeev amplitude obtained
from the integral equation
\begin{equation}
|\phi_{\alpha,n} \rangle = G_0 t  P_\alpha  |\phi_{\alpha,n} \rangle.
\end{equation}

The symmetrized amplitudes  for the two-cluster reactions
are given in Refs.~\cite{deltuva:10c,deltuva:11b,deltuva:07a,deltuva:07c}.
The four-cluster breakup amplitude 
in terms of the symmetrized AGS operators \eqref{eq:U} becomes
\begin{equation} \label{eq:U0}
\begin{split}  
 \langle \Phi_{0} |  T_{0 \alpha} | \Phi_{\alpha,n} \rangle 
= {}& S_{0\alpha}  \langle \Phi_{0} | (1+P_1)
\{ [1+P_{34}(1+P_1)] \\
{}& \times t \, G_0    U_1 G_0 \, t \, G_0 \, \mcu_{1\alpha} \\ {}& + 
(1+P_2) t \, G_0    U_2 G_0 \,  t \, G_0 \, \mcu_{2\alpha} \}
| \phi_{\alpha,n} \rangle .
\end{split}
\end{equation}
The factors $S_{0\alpha}$ depend on the symmetry 
of the final channel state $|\Phi_{0} \rangle $. For example,
$S_{01} = \sqrt{6}$ and $S_{02} = 2$ for
nonsymmetrized  $|\Phi_{0} \rangle $, while
$S_{01} = \sqrt{3}$ and $S_{02} = \sqrt{2}$ for
  $|\Phi_{0} \rangle $  that is symmetrized within the pair (12).

We denote by  $| \Phi_{0}^0 \rangle $ the four-cluster channel state
at threshold, where  $E=0$ and all momenta  vanish. Since all particles have
equal momenta, the channel state $| \Phi_{0}^0 \rangle $ is invariant
under all permutations, i.e., 
$ P_{ab} | \Phi_{0}^0 \rangle = | \Phi_{0}^0 \rangle$.
In this particular case the  relation \eqref{eq:U0} simplifies to
\begin{equation} \label{eq:U0_0}
\begin{split}  
 \langle \Phi_{0}^0 |  T_{0 \alpha} | \Phi_{\alpha,n} \rangle 
= {}& S_{0\alpha}  \langle \Phi_{0}^0 | 
 (12 \, t \, G_0    U_1 G_0 \, t \, G_0 \, \mcu_{1\alpha} \\ {}& + 
6 \, t \, G_0    U_2 G_0 \, t \, G_0 \, \mcu_{2\alpha} )
| \phi_{\alpha,n} \rangle .
\end{split}
\end{equation}

We use the momentum-space partial-wave  framework to
solve the AGS equations \eqref{eq:U}. Two different types of basis states
$|k_x k_y k_z [(l_x l_y)J l_z] \mathcal{JM} \rangle_\alpha =
|k_x k_y k_z \nu \rangle_\alpha $ with
$\alpha = 1$ and 2 are employed.
Here $k_x$, $k_y$, and $k_z$ denote  magnitudes of the Jacobi momenta.
For $\alpha=1$ the Jacobi momenta describe the relative motion in the 1+1, 
2+1, and 3+1 subsystems and are expressed in terms of single particle
momenta $\mbf{k}_{a}$ as
\begin{subequations} \label{eq:jacobi1}
\begin{align}  
\mbf{k}_x = {}& \frac12(\mbf{k}_2 -\mbf{k}_1), \\
\mbf{k}_y = {}& \frac13[2\mbf{k}_3 -(\mbf{k}_1+\mbf{k}_2)], \\
\mbf{k}_z = {}& \frac14[3\mbf{k}_4 -(\mbf{k}_1+\mbf{k}_2+\mbf{k}_3)],
\end{align}
\end{subequations}
while for $\alpha=2$ they describe the relative motion in the 1+1, 1+1, and 
2+2 subsystems, i.e.,
\begin{subequations} \label{eq:jacobi2}
\begin{align}  
\mbf{k}_x = {}& \frac12(\mbf{k}_2 -\mbf{k}_1), \\
\mbf{k}_y = {}& \frac12(\mbf{k}_4 -\mbf{k}_3), \\
\mbf{k}_z = {}& \frac12[(\mbf{k}_4+\mbf{k}_3) -(\mbf{k}_1+\mbf{k}_2)].
\end{align}
\end{subequations}
The respective orbital angular momenta $l_x$, $l_y$, and $ l_z$ 
are coupled to the total angular momentum  $\mathcal{J}$ with the
projection  $\mathcal{M}$; all discrete quantum numbers are abbreviated
by $\nu$. An explicit form of integral equations is obtained by 
inserting the respective completeness relations
\begin{equation} \label{eq:k1k}
1 = \sum_{\nu} \int_0^\infty |k_x k_y k_z \nu \rangle_\alpha
k_x^2 dk_x \, k_y^2 dk_y \, k_z^2 dk_z \,
{}_{\alpha}\langle k_x k_y k_z \nu | 
\end{equation}
between all operators in Eqs.~\eqref{eq:U}. Due to rotational symmetry all 
operators are diagonal in $\mathcal{J}$ and independent of $\mathcal{M}$.
 In addition, $U_\alpha$ are diagonal
in $k_z$, $J$ and $l_z$, $P_{34}$ is diagonal in $k_x$ and $l_x$,
$t$ is diagonal in  $k_y$, $k_z$ and all $\nu$, and $G_0$ is diagonal
in all quantum numbers.
In this representation  the  AGS equations for each $\mathcal{J}$
become a system of coupled  integral equations in three continuous variables
 $k_x$, $k_y$, and $k_z$. Such three-variable equations
have been solved in Refs.~\cite{deltuva:07a,deltuva:07c}
for the four-nucleon scattering below the three-cluster breakup threshold.
The kernel of integral equations \eqref{eq:U} contains integrable
singularities arising from each subsystem bound state pole of $U_\alpha$;
we isolate these poles  in different subintervals and treat them by the
subtraction technique when  integrating over $k_z$ \cite{deltuva:07a}.
However, in the energy regime of the four-cluster breakup,
$E \ge 0$,  the AGS equations  \eqref{eq:U} contain additional
singularities arising from $G_0$, namely,
${}_{\alpha}\langle k_x' k_y' k_z'\nu' | G_0 | k_x k_y k_z \nu \rangle_{\alpha} = 
\delta_{\nu'\nu} \, \delta(k_x'-k_x) \delta(k_y'-k_y) \delta(k_z'-k_z) / 
[k_x^2 k_y^2 k_z^2  
(E+i0-k_x^2/2\mu_{\alpha x}- k_y^2/2\mu_{\alpha y} - k_z^2/2\mu_\alpha)]$,
where $\mu_{\alpha x}$  and $\mu_{\alpha y}$ 
are the respective reduced masses. 
The operator products like $P_{34} P_{1} G_0$
render the singularity structure of Eqs.~\eqref{eq:U} very complicated.
It is considerably simpler at $E=0$ where there is only one singular point
$ k_x = k_y = k_z = 0$. In this case  the integrals involving $G_0$  over 
any of the Jacobi momenta $k_j$ 
(except for the trivial integrals involving momentum $\delta$-functions) 
are of type
\begin{equation} \label{eq:g0}
\int_0^\infty \frac{f(k_x,k_y,k_z) k_j^2 dk_j}{i0- 
k_x^2/2\mu_{\alpha x}- k_y^2/2\mu_{\alpha y} - k_z^2/2\mu_\alpha}
\end{equation}
where the function $f(k_x,k_y,k_z)$ is regular at $k_j=0$. Thus, the 
the singularity of $G_0$ is cancelled by $k_j^2$ making
the integral numerically harmless.
Nevertheless, $G_0$ varies very rapidly near $ k_x = k_y = k_z = 0$
and therefore one should avoid interpolation of $G_0$; this will be
explained in Sec.~\ref{sec:sep}.

The four free boson channel state $|\Phi_0^0 \rangle$ has
$ k_x = k_y = k_z = 0$ and $l_x=l_y=l_z=J=\mathcal{J} = 0$.
Due to the threshold law the only nonvanishing partial-wave component 
of the breakup amplitude \eqref{eq:U0_0} is 
the one  with $\mathcal{J} = 0$, i.e.,
\begin{equation}
 \langle \Phi_{0}^0 |  T_{0 \alpha}| \Phi_{\alpha,n} \rangle = 
\langle \Phi_{0}^0 |T_{0 \alpha}^{\mathcal{J}=0}|\Phi_{\alpha,n} \rangle /(4\pi)^2.
\end{equation}
Although $|\Phi_0^0 \rangle$  has only the $l_j = J=\mathcal{J}=0$ component, 
we emphasize that angular momenta $l_j$ and $J$ are not conserved.
Nonzero $l_j$ and $J$ basis states  must be included  
when solving the AGS equations \eqref{eq:U}.

\subsection{Separable potential \label{sec:sep}}

The universal properties of the four-boson system  must be independent 
of the short-range interaction details. 
We choose the two-boson potential of a separable form
acting in the $S$-wave only, i.e.,
\begin{equation} \label{eq:vsep}
v = |g\rangle \lambda \langle g| \, \delta_{l_x 0}.
\end{equation}
The form factor 
\begin{equation} \label{eq:gsep}
\langle k_x |g\rangle = [1+c_2\,(k_x/\Lambda)^2]e^{-(k_x/\Lambda)^2}
\end{equation}
is taken over from  Ref.~\cite{deltuva:10c}.
Two very different choices of $c_2$, namely, $c_2 = 0$ and $c_2 = -9.17$,
are used in order to confirm the universality of the obtained results.
 The strength 
\begin{equation} \label{eq:lsep}
\lambda = \frac{2}{\pi m}
 \left\{ \frac1a - \left[1+\frac{c_2}{2} \left(1+\frac{3 c_2}{8} \right) \right]
\frac{\Lambda}{\sqrt{2\pi}} \right\}^{-1}
\end{equation}
is constrained to reproduce the given value 
of the scattering length $a$ for two particles of mass $m$.
The potential \eqref{eq:vsep} supports one shallow two-boson bound state at 
$a>0$ and no one at $a<0$. 
There are no deeply bound dimers.

The resulting two-boson transition matrix 
\begin{equation} \label{eq:tsep}
t = |g\rangle \tau \langle g|
\end{equation}
is  separable as well with
 $\tau = (\lambda^{-1}- \langle g|  G_0 | g \rangle)^{-1}$.
This allows to reduce the AGS equations \eqref{eq:U}
to a system of two-variable ($k_y$ and $k_z$) integral equations
and thereby simplifies considerably the numerical calculations.
The AGS equations with separable interactions are solved in the form
\begin{subequations} \label{eq:U-g}
\begin{align}  \nonumber
\langle g|  G_0 \mcu_{11}| \phi_{1,n} \rangle  =
{}&  P_{34} \langle g|  P_1 | \phi_{1,n} \rangle \\  \nonumber
{}& +  P_{34} \langle g| G_0 U_1 G_0 |g\rangle \tau
\langle g| G_0  \mcu_{11} | \phi_{1,n} \rangle \\
{}& +  \langle g|  G_0 U_2 G_0   |g\rangle \tau
\langle g| G_0  \mcu_{21} | \phi_{1,n} \rangle,
\label{eq:K11} \\  \nonumber
\langle g|  G_0 \mcu_{21}| \phi_{1,n} \rangle
= {}&  (1 +  P_{34})   \langle g|  P_1 | \phi_{1,n} \rangle \\  
{}& + (1 +  P_{34})  \langle g|  G_0 U_1 G_0   |g\rangle
\tau \langle g| G_0  \mcu_{11}  | \phi_{1,n} \rangle, \label{eq:K21} \\
\langle g|  G_0 \mcu_{12}| \phi_{2,n} \rangle  = \nonumber
{}&   \langle g|  P_2 | \phi_{2,n} \rangle \\  \nonumber
{}& + P_{34} \langle g| G_0 U_1 G_0 |g\rangle \tau
\langle g| G_0  \mcu_{12} | \phi_{2,n} \rangle \\
{}& +  \langle g|  G_0 U_2 G_0   |g\rangle \tau
\langle g| G_0  \mcu_{22} | \phi_{2,n} \rangle,
\label{eq:K12} \\
\langle g|  G_0 \mcu_{22}| \phi_{2,n} \rangle
= {}&  (1 + P_{34})   \langle g|  G_0 U_1 G_0   |g\rangle
\tau \langle g| G_0  \mcu_{12}  | \phi_{2,n} \rangle. \label{eq:K22}
\end{align}
\end{subequations}
We emphasize that with respect to angular momentum quantum numbers, 
the full complexity of the four-body problem is still present in 
Eqs.~\eqref{eq:U-g}.
Basis states  up to $l_y, l_z, J < 3$ 
are included  to achieve the partial-wave convergence for the solutions
of the AGS equations \eqref{eq:U-g}.
Regarding the continuous variables $k_y$ and $k_z$, the treatment is
taken over from  Ref.~\cite{deltuva:07a}. It requires interpolation
of  $\langle g|  G_0 U_\alpha G_0   |g\rangle $ in initial and final 
momenta $k_y$.
To take care of the $G_0$ singularity at $k_j=0$ we factorize 
$\langle g k'_y k'_z \nu' |  G_0 U_\alpha G_0   |g k_y k_z \nu \rangle 
= ({k'_y}^2/2\mu_{\alpha y} + {k'_z}^2/2\mu_\alpha)^{-1} \,
\bar{U}_\alpha^{\nu'\nu} (k'_y,k_y,k_z) \, 
(k_y^2/2\mu_{\alpha y} + k_z^2/2\mu_\alpha)^{-1} \, 
\delta(k'_z-k_z)/k_z^2$ such that 
$\bar{U}_\alpha^{\nu'\nu} (k'_y,k_y,k_z)$ is a very smooth function
and is interpolated very reliably using spline functions
as in Ref.~\cite{deltuva:07a}, while the factors of the type
$(k_y^2/2\mu_{\alpha y} + k_z^2/2\mu_\alpha)^{-1}$ are calculated explicitly
wherever needed.

The discretization of integrals in Eqs.~\eqref{eq:U-g} 
using Gaussian quadrature rules  leads  to a 
system of linear algebraic equations. We reduce its dimension even further 
by applying the so-called $\mathcal{K}$-matrix technique
\cite{deltuva:ef} converting the complex equations into the real ones,
and by eliminating the $\alpha=2$ component, i.e., by
inserting  Eq.~\eqref{eq:K21} into \eqref{eq:K11}
and Eq.~\eqref{eq:K22} into \eqref{eq:K12}.
To avoid difficulties due to possibly slow convergence of iterative methods,
the resulting linear system is solved using the direct matrix inversion.
Further details of the calculations
can be found in Refs.~\cite{deltuva:07a,deltuva:ef}.

\section{Four-boson recombination \label{sec:res}}

The number of the four-atom recombination events 
in a non-degenerated atomic gas 
per volume and time is $K_4 \rho^4$  where $\rho$ is the density of atoms
and $K_4$ the four-atom recombination rate.
In the ultracold limit the kinetic energies of initial atoms are
much smaller than the final two-cluster  kinetic or binding
energies, and the initial momenta are much smaller than the
momenta of the two resulting clusters,
$k_x,k_y,k_z << p_{\alpha, n'}$. Under these conditions
$K_4$ can be approximated very well by the zero-temperature limit
$K_4^0$ calculated at $E=0$ with $k_x=k_y=k_z=0$ in the initial state, i.e.,
\begin{equation} \label{eq:k4}
K_4^0 = 16 \pi^7 \sum_{\alpha,n'}  \mu_{\alpha}  p_{\alpha, n'}
| \langle \Phi_{0}^0 |  T_{0 \alpha}^{\mathcal{J}=0} | \Phi_{\alpha,n'} \rangle |^2.
\end{equation}
It has contributions from all two-cluster channels $(\alpha,n')$.

\begin{figure}[!]
\includegraphics[scale=0.58]{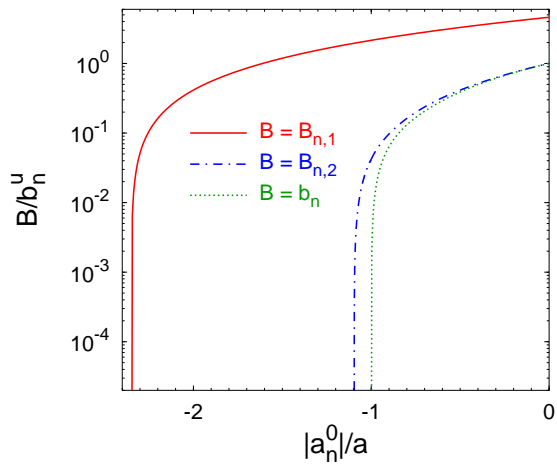}
\caption{\label{fig:B} (Color online)
Trimer and tetramer binding energies as functions of 
the two-boson scattering length.}
\end{figure}

Concerning the four-atom recombination, the most interesting regime 
lies at large negative two-boson scattering length $a$ where
no shallow dimers exist  but the Efimov trimers and tetramers 
cross the zero-energy threshold.
In our nomenclature  we characterize the trimers by one integer number
$n$, starting with $n=0$ for the ground state, and
the tetramers by two integers $(n,k)$,
where $n$ refers to the associated trimer and $k=1$ (2) for a deeper 
(shallower) tetramer. 
The $a$-dependence of the  $n$th  family trimer ($b_n$) and
tetramer ($B_{n,k}$) binding energies is shown in Fig.~\ref{fig:B} for $a<0$.
We denote by $a_n^0$  the specific negative value of $a$ 
where  the $n$th trimer binding energy vanishes, i.e.,  $b_n = 0$.
In the universal limit, i.e., for sufficiently large $n$,
the ratio $a_{n+1}^0/a_n^0 = e^{\pi/s_0} \approx 22.6944$
where $s_0 \approx 1.00624$ \cite{braaten:rev}.
Trimer binding energy $b_n^u$ taken in the unitary limit $1/a = 0$,
is used to build dimensionless ratios $b_n/b_n^u$ and
$B_{n,k}/b_n^u$. These ratios become 
independent of the short-range details of the interaction 
provided that its range is small enough  compared to 
the size of the few-boson states. This condition is fulfilled
for high excited states, i.e., for sufficiently large $n$.
We checked this independence of the results in Fig.~\ref{fig:B}
performing calculations with $n=3$ and 4 
for two choices of the potential form factor \eqref{eq:gsep}.

We remind that only two lowest $n=0$ tetramers are true bound states.
All higher tetramers lie above the lowest atom-trimer threshold
and therefore are unstable bound states with finite width and lifetime.
Their positions $B_{n,k}$ shown in Fig.~\ref{fig:B} are extracted from 
the atom-trimer scattering
calculations at $E<0$ as described in Refs.~\cite{deltuva:10c,deltuva:11a}.
The specific negative values of $a = a_{n,k}^0$ 
where  the respective tetramers emerge at the four free particle threshold
can be obtained by extrapolating the results in
Fig.~\ref{fig:B} to  $B_{n,k}=0$. This procedure yields
universal ratios 
\begin{subequations} \label{eq:a-0}
\begin{align}
a_{n,1}^{0}/ a_{n}^{0} & =  0.4254(2), \\
a_{n,2}^{0}/ a_{n}^{0} & =  0.9125(2).
\end{align}
\end{subequations}
The uncertainties, apart from the numerical accuracy, arise due
to the residual dependence on $n \ge 3$ and $c_2$ in Eq.~\eqref{eq:gsep}.

\begin{figure}[!]
\includegraphics[scale=0.64]{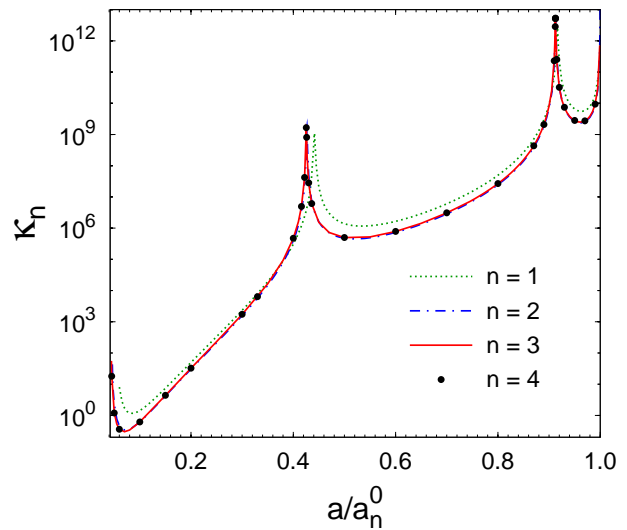}
\caption{\label{fig:k4} (Color online)
Dimensionless four-boson recombination rate $\kappa_n$
as a function of two-boson scattering length.}
\end{figure}

We will show our  results for the four-boson recombination rate
as  functions of the dimensionless ratio $a/a_n^0$. Then in the
interval $(e^{-\pi/s_0},1)$ there are exactly $n$ trimer states
with $0 \le n' \le n-1$ that contribute to Eq.~\eqref{eq:k4}.
Instead of $K_4^0$ we build a dimensionless quantity
\begin{equation} \label{eq:kp4}
\kappa_n = K_4^0 m/(\hbar |a_n^0|^7).
\end{equation}
With $1 \le n \le 4$ in Fig.~\ref{fig:k4}
we explore a broad range of the
two-boson scattering length $a$ and recombination channels $n'$
whereas $\alpha=1$.
The four-boson recombination rate varies over many orders of magnitude,
but all $\kappa_n$ as functions of the respective $a/a_n^0$
show qualitatively the same behavior. For higher $n$, i.e., $n\ge 3$, 
there is also a very good quantitative agreement, indicating
the universality of our $\kappa_n$ results. Indeed, while $\kappa_n$
 in Fig.~\ref{fig:k4} are obtained with $c_2=0$ in Eq.~\eqref{eq:gsep},
additional calculations with $c_2=-9.17$ for $n=3$ and 4
agree very well with the corresponding predictions in  Fig.~\ref{fig:k4},
thereby confirming our conclusion on $\kappa_n$ universality.
In contrast, $\kappa_1$ shows significant quantitative deviations
from the universal behavior  due to finite-range effects: 
in the regime  $a/a_1^0 < 1$ the only available recombination channel
leads to the ground state trimer whose size is comparable to
the interaction range \cite{lazauskas:he}. 
We note that our non-universal results at $n\le 1$ 
depend not only on  $c_2$ but also on $m$ and $\Lambda$; 
our calculations are performed with $m$ being the mass of ${}^4$He atom
and $\Lambda = 0.4$ {\AA}${}^{-1}$.

The  four-boson recombination rate has resonant peaks at
$a = a_{n,k}^0$ where the respective tetramers are at the
four free atom threshold. The ratios  $a_{n,k}^0/a_n^0$ extracted
from the peak positions 
(from $B_{n,k}=0$ in the case $n=0$) are collected in Table~\ref{tab:a}.
For  $n\ge 3$ they approach the universal limit with high accuracy
and are fully consistent with the ones in Eqs.~\eqref{eq:a-0}.
Furthermore, the four-boson recombination rate has sharp cusps at 
$a = a_{n}^0$, or, equivalently, at $a = e^{-\pi/s_0}a_{n}^0$, where
a new atom-trimer channel opens. We list in Table~\ref{tab:a}
also the ratios  $a_{n+1}^0/a_n^0$ to prove that our numerical predictions
converge towards analytical result $e^{\pi/s_0} \approx 22.6944$
\cite{efimov:plb,braaten:rev}.
Furthermore, for $n\ge 3$ we obtain  
$a_{n}^{0} \sqrt{m b_n^u}/\hbar = -1.5077(1)$.

\begin{table}[!]
\begin{ruledtabular}
\begin{tabular}{*{4}{l}} $n$ & 
$ a_{n,1}^{0}/ a_{n}^{0}$ & $ a_{n,2}^{0}/ a_{n}^{0}$ & $ a_{n+1}^{0}/ a_{n}^{0}$
\\  \hline
0 & 0.4435 & 0.8841 & 17.752 \\
1 & 0.4412 & 0.9162 & 21.935 \\
2 & 0.4267 & 0.9128 & 22.639 \\
3 & 0.4254 & 0.9124 & 22.691 \\
4 & 0.4254 & 0.9125 & 22.694 \\
\hline
3 & 0.4253 & 0.9125 & 22.695 \\
4 & 0.4254 & 0.9125 & 22.694 \\
\end{tabular}
\end{ruledtabular}
\caption{ \label{tab:a}
Ratios of special two-boson scattering length values
corresponding to the peaks and cusps of the four-boson recombination rate
($n\ge 1$) or vanishing tetramer binding energy ($n=0$).
Results are obtained using potential form factor with $c_2 = 0$ (top) 
and $c_2=-9.17$ (bottom).}
\end{table}

\begin{figure}[!]
\includegraphics[scale=0.64]{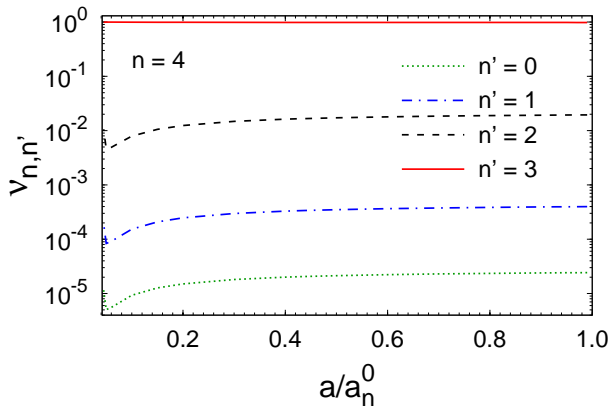}
\caption{\label{fig:nu} (Color online)
Relative weights $\nu_{n,n'}$ for the production of the
$n'$th state trimers via four-boson recombination.
$n=4$ atom-trimer channels contribute in the shown regime.}
\end{figure}

At $a/a_1^0>1$  not only the ground state but also excited trimers are 
produced via four-atom recombination and contribute to the total
rate \eqref{eq:k4}. In Fig.~\ref{fig:nu} we compare the
relative weights $\nu_{n,n'}$ for the production of the
trimers in the individual states $n'$ when $n$ 
atom-trimer channels are available in total.
The weights are normalized to $\sum_{n'=0}^{n-1}\nu_{n,n'} = 1$. 
The results in Fig.~\ref{fig:nu} clearly demonstrate that
more weakly bound trimers are produced more efficiently and
 the four-atom recombination process is strongly dominated
by the $n'=(n-1)$th channel with the shallowest available trimer.
We note that a similar conclusion was drawn in Ref.~\cite{deltuva:11b}
about the trimer production in dimer-dimer collisions.
Except for the vicinity of  $a/a_{n}^0 = e^{-\pi/s_0}$
where the shallowest trimer disappears, the 
relative weights $\nu_{n,n'}$ depend quite weakly on $a$.
Thus, the tetramer states around $a = a_{n,k}^0$
enhance the production of all trimers by equal factors.

As mentioned in the introduction, the recombination of four identical 
bosons has been calculated in Refs.~\cite{stecher:09a,mehta:09a}
using the adiabatic hyperspherical framework.
There is a good  qualitative agreement between our results and those of
Refs.~\cite{stecher:09a,mehta:09a}: 
the shape of the recombination rate, including
its resonant peaks and cusps, is reproduced in both cases.
However, the calculations of Refs.~\cite{stecher:09a,mehta:09a}
are limited to $n=1$  where the finite-range effects are
not entirely negligible and deviations from our
universal results can be expected much like in the case of our
$n=1$ results in Fig.~\ref{fig:k4}. Within this accuracy
the predictions of Refs.~\cite{stecher:09a,mehta:09a} for the
 peak positions, i.e.,
$ a_{n,1}^{0}/a_{n}^{0} \approx 0.43$ and $ a_{n,2}^{0}/a_{n}^{0} \approx 0.90$,
are consistent with our results.
On the other hand, the available experimental data \cite{ferlaino:09a} 
are obtained in the system of ultracold ${}^{133}$Cs atoms with deeply bound  
non-Efimov-like few-atom states and refer to $n=0$. 
Therefore it is not surprising that the experimental results 
$a_{n,1}^{0}/a_{n}^{0} \approx 0.47$ and $ a_{n,2}^{0}/a_{n}^{0} \approx 0.84$
deviate from the universal values even more. Nevertheless, they
remain roughly consistent with the theoretical predictions. 
Thus, our calculations present no improvement over the ones
of Refs.~\cite{stecher:09a,mehta:09a} when comparing with the
existing data, but serve as a theoretical benchmark and may be valuable 
for the future experiments performed in a truly universal regime.

\section{Summary \label{sec:sum}}

We studied ultracold four-boson recombination. 
Exact  four-particle scattering equations  in the AGS version have been used.
We related the four-cluster breakup and recombination operators 
to the standard two-cluster AGS operators.
The scattering equations at the four-cluster breakup threshold
have been precisely solved  in the momentum-space  framework.
The zero-temperature limit of the
four-boson recombination rate was calculated for negative values
of the two-boson scattering length. 
We demonstrated its universal behavior
in the regime with high excited Efimov trimers where the 
finite-range effects are negligible, and determined the positions
of resonant recombination peaks caused by the tetramer states.
We also have shown that most of the Efimov trimers produced via the 
four-boson recombination are in the state with weakest binding.
Our results are consistent with existing experimental data \cite{ferlaino:09a}
and theoretical predictions \cite{stecher:09a,mehta:09a}, but
exceed the latter ones with respect to the accuracy in the universal limit.
In addition, the present work constitutes a step 
towards solving the momentum-space four-body scattering equations 
above the breakup threshold.

\begin{acknowledgments}
The author thanks  P.~U.~Sauer for discussions.
on few-body recombination problem.
\end{acknowledgments}


\end{document}